\documentclass[10pt]{article}
\usepackage{amsfonts}
\usepackage{amsmath}
\usepackage{amssymb}
\usepackage{mathptmx}
\usepackage{geometry}
\usepackage{graphicx}
\usepackage{hyperref}
\usepackage{cancel}
\usepackage{textcomp}
\usepackage{bm}
\usepackage{times}
\usepackage{epsfig}
\usepackage{xcolor}
\usepackage{slashed}
\usepackage{booktabs}
\usepackage{latexsym}
\usepackage{verbatim}
\usepackage{extarrows}
\usepackage{multirow}

\newcommand{\Tr}{\rm Tr}

\newcommand{\MS}{\rm MS}

\newcommand{\GeV}{\rm GeV}

\newcommand{\calO}{{\cal O}}

\begin{document}
\baselineskip=16pt

\pagenumbering{arabic}

\vspace{1.0cm}

\begin{center}
{\Large\sf Renormalization group evolution of dimension-seven baryon- and lepton-number-violating operators}
\\[10pt]
\vspace{.5 cm}

{Yi Liao~$^{a,b,c}$\footnote{liaoy@nankai.edu.cn} and Xiao-Dong Ma~$^{a}$\footnote{maxid@mail.nankai.edu.cn}}

{
$^a$~School of Physics, Nankai University, Tianjin 300071, China
\\
$^b$ CAS Key Laboratory of Theoretical Physics, Institute of Theoretical Physics,
Chinese Academy of Sciences, Beijing 100190, China
\\
$^c$ Center for High Energy Physics, Peking University, Beijing 100871, China}

\vspace{2.0ex}

{\bf Abstract}

\end{center}

We study dimension-seven operators in standard model effective field theory. These operators are classified into two sets, one violating lepton number but preserving baryon number ($\Delta L=\pm 2$, $\Delta B=0$) and the other violating both but preserving their sum ($-\Delta L=\Delta B=\pm 1$). It has been found in the previous literature that there are respectively 13 and 7 such independent operators. We show that one operator is redundant in each set so that the complete list contains only 12 and 6 operators respectively. We accomplish this by using standard model equations of motion and various Fierz identities. We calculate the one-loop anomalous dimension matrix for the 6 operators in the second set, and illustrate its possible phenomenological implications by working out renormalization group running of the Wilson coefficients that could contribute to the type of proton decays with $-\Delta L=\Delta B=\pm 1$, such as $p\to\nu\pi^+$.

\newpage

\section{Introduction}

When the standard model (SM) is regarded as an effective field theory, the low energy effects from physics beyond SM at a high energy scale can be organized in a tower of high dimensional operators. These operators are composed of the SM fields, respect the SM gauge symmetry $SU(3)_C\times SU(2)_L\times U(1)_Y$, and have coefficients that are suppressed by relevant high energy scales. This forms what is usually called standard model effective field theory (SMEFT) \cite{Georgi:1991ch,Georgi:1994qn,Manohar:1996cq,Kaplan:2005es,Burgess:2007pt,Skiba:2010xn}. The accidental symmetries in SM such as lepton and baryon number conservation are generically not preserved any longer by high dimensional operators; this is not surprising if one imagines that a high scale supersymmetric or grand unification theory is responsible for those operators.

The tower of high dimensional operators starts at dimension five, and it turns out that there is a unique operator \cite{Weinberg:1979sa}, which violates lepton number by two units ($\Delta L=\pm 2$) and can accommodate a Majorana mass for neutrinos. There are much more operators at dimension six \cite{Buchmuller:1985jz}, and the complete list contains 63 independent operators \cite{Grzadkowski:2010es}, among which 59 conserve both lepton and baryon number and the other 4 violate both by one unit ($\Delta L=\Delta B=\pm 1$). The latter can induce nucleon decays such as $p\to e^+\pi^0,~\bar\nu\pi^+$, and $n\to\bar\nu\pi^0$. The first systematic analysis on dimension seven (dim-7) operators has been made recently in \cite{Lehman:2014jma}; for earlier studies, see \cite{Weinberg:1980bf,Weldon:1980gi}, and for a recent analysis of dim-7 operators in SM extended by right-handed neutrinos, see \cite{Bhattacharya:2015vja}. It is found that there are altogether 20 operators, 13 of which violate lepton number but preserve baryon number ($\Delta L=2$, $\Delta B=0$) and 7 of which violate both but preserve their sum ($-\Delta L=\Delta B=1$). The first set includes the dim-7 generalization of the dim-5 Majorana neutrino mass operator which turns out to be also unique, consistent with the general analysis in \cite{Liao:2010ku}. The second set could induce another type of rare nucleon decays such as $p\to\nu\pi^+$ and $n\to e^-\pi^+,~\nu\pi^0$. The pursuit of high dimensional operators can be continued. For instance, at dimension 9, operators that violate baryon number by two units start to appear. These operators could induce phenomena such as neutron-antineutron oscillation, and may bridge our understanding of some underlying theory and the matter-antimatter asymmetry observed in the Universe \cite{Mohapatra:2009wp}. For a recent discussion on the relation between the dimension and lepton/baryon number of operators, see Ref.~\cite{Kobach:2016ami}.

The high dimensional operators discussed above are generated at a high energy scale where one integrates out certain heavy degrees of freedom from an underlying theory. When studying the physical effects of those operators in low energy experiments, it is necessary to run them down to the scale at which their matrix elements are evaluated. This can be accomplished by renormalization group equations (RGEs), and it boils down to the computation of the anomalous dimension matrix for relevant operators. For the unique dim-5 operator the one-loop analysis has been carried out previously in Ref. \cite{Babu:1993qv} with the final answer reported in \cite{Antusch:2001ck}. The situation becomes complicated for dim-6 operators as there are too many of them and strong interactions also set in. The computation of the anomalous dimension matrix has been recently accomplished in a series of papers \cite{Grojean:2013kd,Elias-Miro:2013gya,Elias-Miro:2013mua,Jenkins:2013zja,Jenkins:2013wua,Alonso:2013hga,
Alonso:2014zka}. The purpose of our current work is to initiate the evaluation of the one-loop anomalous dimension matrix for dim-7 operators. Since, as mentioned above, these operators fall into two sets according to whether baryon number is conserved or not, the two sets do not mix under one-loop renormalization. We report our result in this work on the second set of operators that violate baryon number conservation and defer the discussion on the first set of operators in a future publication. In so doing, we also find that the basis of operators established in \cite{Lehman:2014jma} can be further shortened, with one less operator in each set.

This paper is organized as follows. In Sec. \ref{pre} we set up our conventions and show the redundancy of the basis given in \cite{Lehman:2014jma} by establishing two linear relations that can be used to reduce one operator in each of the two sets of operators. We compare our count of independent operators with those in the literature. Then in Sec. \ref{rge}, we present our result on the one-loop RGEs for the Wilson coefficients of the 6 dim-7 operators in the second set. We discuss briefly its possible implications on proton decays that violate both baryon and lepton numbers but conserve their sum, such as $p\to \nu \pi^+$. We recapitulate our results in Sec.\ref{conclusion}. Some useful Fierz identities employed in Secs. \ref{pre} and \ref{rge} are collected in the Appendix.

\section{Basis of operators}
\label{pre}

We start with some preliminary discussions. $L,~Q$ are the SM left-handed lepton and quark doublet fields, $u,~d,~e$ are the right-handed up-type quark, down-type quark and charged lepton singlet fields, and $H$ denotes the Higgs doublet. Dropping gauge-fixing terms, the SM Lagrangian is
\begin{eqnarray}
\label{sml}
\nonumber
\mathcal{L}_4&=&
-\frac{1}{4}G^A_{\mu\nu}G^{A\mu\nu}-\frac{1}{4}W^I_{\mu\nu}W^{I\mu\nu} -\frac{1}{4}B_{\mu\nu}B^{\mu\nu}+(D_\mu H)^\dagger(D^\mu H)
-\lambda\left(H^\dagger H-\frac{1}{2}v^2\right)^2
\\
&&+\sum_{\Psi=Q, L, u, d, e}\bar{\Psi}i \slashed{D}\Psi
-\left[\bar{Q}Y_u u \tilde{H}+\bar{Q}Y_d d H+\bar{L}Y_e e H +\mbox{h.c.}\right].
\end{eqnarray}
Here the superscripts $A$ and $I$ count the generators of the $SU(3)_C$ and $SU(2)_L$ group, respectively, $Y_u,~Y_d,~Y_e$ are the Yukawa couplings which are complex matrices in flavor space, and $\tilde H_i=\epsilon_{ij}H^*_j$. The covariant derivative is defined by
\begin{eqnarray}
D_\mu=\partial_\mu-ig_3 T^AG^A_\mu-ig_2T^IW^I_\mu-ig_1YB_\mu,
\end{eqnarray}
where $T^A,~T^I,~Y$ are the generator matrices appropriate for the fields to be acted on. From Eq. (\ref{sml}) one can derive the following equations of motion (EoMs) which will be used to remove redundant operators at one-loop level,
\begin{eqnarray}
i\slashed{D}L&=&Y_e e H,
\\
i\slashed{D}d&=&Y^\dagger_d {H}^\dagger Q,
\end{eqnarray}
or more explicitly in flavor indices,
\begin{eqnarray}
\label{eom1}
i\gamma^\mu D_\mu L^i_t&=&(Y_e)_{tu} e_u H^i,
\\
\label{eom2}
i\gamma_\mu D^\mu_{\sigma \rho}d_{\rho t}&=&(Y^\dagger_d)_{tu}\delta_{kl}Q_{k \sigma u} {H}^*_l.
\end{eqnarray}
We use the letters $p,r,s,t,u,v,w$ for flavors, $i,j,k,l$ and $\alpha,\beta,\sigma,\rho$ for indices in the fundamental representations of $SU(2)_L$ and $SU(3)_C$ respectively. A repeated index is always implied to be summed over.

\begin{table}
\centering
\begin{tabular}{|c|c|c|c|}
 \multicolumn{2}{c}{$\psi^2H^4+\mbox{h.c.}$} &  \multicolumn{2}{c}{ $\psi^2H^3D+\mbox{h.c.}$}
\\
\hline
$\mathcal{O}_{LH}$ & $\epsilon_{ij}\epsilon_{mn}(L^iCL^m)H^jH^n(H^\dagger H)$ & $\mathcal{O}_{LeHD}$ & $\epsilon_{ij}\epsilon_{mn}(L^iC\gamma_\mu e)H^jH^miD^\mu H^n$
\\
\hline
 \multicolumn{2}{c}{$\psi^2H^2D^2+\mbox{h.c.}$}&  \multicolumn{2}{c}{$\psi^2H^2X+\mbox{h.c.}$}
\\
\hline
$\mathcal{O}_{LHD1}$ &  $\epsilon_{ij}\epsilon_{mn}(L^iCD^\mu L^j)H^m(D_\mu H^n)$ &$\mathcal{O}_{LHB}$    &$\epsilon_{ij}\epsilon_{mn}(L^iCi\sigma_{\mu\nu}L^m)H^jH^nB^{\mu\nu}$ \\
$\mathcal{O}_{LHD2}$  & $\epsilon_{im}\epsilon_{jn}(L^iCD^\mu L^j)H^m(D_\mu H^n)$ & $\mathcal{O}_{LHW}$  &$\epsilon_{ij}(\tau^I\epsilon)_{mn}(L^iCi\sigma_{\mu\nu}L^m)H^jH^nW^{I\mu\nu}$ \\
\hline
  \multicolumn{2}{c}{$\psi^4D+\mbox{h.c.}$}  &   \multicolumn{2}{c}{$\psi^4H+\mbox{h.c.}$}\\ \hline
$\mathcal{O}_{\bar{d}uLLD}$ & $\epsilon_{ij}(\bar{d}\gamma_\mu u)(L^iCiD^\mu L^j)$ & $\mathcal{O}_{\bar{e}LLLH}$ & $\epsilon_{ij}\epsilon_{mn}(\bar{e}L^i)(L^jCL^m)H^n$\\
$\mathcal{O}_{\bar{L}QddD}$ & $(\bar{L}\gamma_\mu Q)(dCiD^\mu d)$ & $\mathcal{O}_{\bar{d}LQLH1}$ & $\epsilon_{ij}\epsilon_{mn}(\bar{d}L^i)(Q^jCL^m)H^n$\\
$\mathcal{O}_{\bar{e}dddD}$  & $(\bar{e}\gamma_\mu d)(dCiD^\mu d)$ & $\mathcal{O}_{\bar{d}LQLH2}$ & $\epsilon_{im}\epsilon_{jn}(\bar{d}L^i)(Q^jCL^m)H^n$\\
   &  & $\mathcal{O}_{\bar{d}LueH}$ & $\epsilon_{ij}(\bar{d}L^i)(uCe)H^j$\\
   &  & $\mathcal{O}_{\bar{Q}uLLH}$ & $\epsilon_{ij}(\bar{Q}u)(LCL^i)H^j$\\
   &  & $\mathcal{O}_{\bar{L}dud\tilde{H}}$ & $(\bar{L}d)(uCd)\tilde{H}$\\
   &  & $\mathcal{O}_{\bar{L}dddH}$ & $(\bar{L}d)(dCd)H$\\
   &  & $\mathcal{O}_{\bar{e}Qdd\tilde{H}}$ & $\epsilon_{ij}(\bar{e}Q^{i})(dCd)\tilde{H}^j$\\
   &  & $\mathcal{O}_{\bar{L}dQQ\tilde{H}}$ & $\epsilon_{ij}(\bar{L}d)(QCQ^{i})\tilde{H}^j$ \\ \hline
 \multicolumn{4}{c}{redundant operators} \\
 \hline
$\mathcal{O}^{(2)}_{\bar{d}uLLD}$&  $\epsilon_{ij}(\bar{d}\gamma_\mu u)(L^iC\sigma^{\mu\nu}D_\nu L^j)$ & $\mathcal{O}_{\bar{L}dQdD}$ & $(\bar{L}iD^\mu  d)(QC\gamma_\mu d)$ \\ \hline
\end{tabular}
\caption{The basis of the twenty dim-7 operators in Ref. \cite{Lehman:2014jma} is reproduced here with some modifications. The flavor and summed color indices are not shown. (1) We label operators in a more symmetric manner. (2) We associate a factor of $i$ with the gauge covariant derivative $D_\mu$ and the matrix $\sigma_{\mu\nu}$ (but drop $i^2$ from the $\Psi^2H^2D^2$ operators) for convenience of later RGE analysis. (3) We replace the original operator $\epsilon_{ij}(L^iC\gamma_\mu e)(\bar{d}\gamma^\mu u)H^j$ by the new one, $\mathcal{O}_{\bar{d}LueH}$, so that all operators in the $\Psi^4H$ sector are products of scalar bilinears. (4) The two redundant operators listed in the last row are to be removed. The equivalence of the two operators in (3) and redundancy in (4) are established in the main text. }
\label{tab1}
\end{table}

The twenty dim-7 operators listed in Ref. \cite{Lehman:2014jma} are shown in Table \ref{tab1} with some modifications. Our notations for operators are such that the fermion fields and their flavors are identically ordered and follow the chains of the two bilinears involved. For instance, the six independent and complete operators in the second set that violate both baryon and lepton numbers but preserve their sum, i.e., $-\Delta L=\Delta B=1$, are written as,
\begin{eqnarray}
\label{eq_BL}
\nonumber
\mathcal{O}^{prst}_{\bar{L}dud\tilde{H}}&=&
\epsilon_{\alpha \beta\sigma}\epsilon_{ij}(\bar{L}_{ip}d_{\alpha r})(u_{\beta s}Cd_{\sigma t})H^*_j,
\\
\nonumber
\mathcal{O}^{prst}_{\bar{L}dddH}&=&
\epsilon_{\alpha \beta\sigma}\delta_{ij}(\bar{L}_{ip}d_{\alpha r})(d_{\beta s}Cd_{\sigma t})H_j,
\\
\nonumber
\mathcal{O}^{prst}_{\bar{e}Qdd\tilde{H}}&=&
-\epsilon_{\alpha \beta\sigma}\delta_{ij}(\bar{e}_{p}Q_{i\alpha r})(d_{\beta s}Cd_{\sigma t})H^*_j,
\\
\nonumber
\mathcal{O}^{prst}_{\bar{L}dQQ\tilde{H}}&=&
-\epsilon_{\alpha \beta\sigma}\delta_{kl}\delta_{ij}(\bar{L}_{kp}d_{\alpha r})
(Q_{l\beta s}CQ_{i\sigma t})H^*_j,
\\
\nonumber
\mathcal{O}^{prst}_{\bar{L}QddD}&=&
\epsilon_{\alpha \beta\sigma}\delta_{ij}(\bar{L}_{ip}\gamma_\mu Q_{j\alpha r})
(d_{\beta s}CiD^\mu_{\sigma \rho} d_{\rho t}),
\\
\mathcal{O}^{prst}_{\bar{e}dddD}&=&
\epsilon_{\alpha \beta\sigma}(\bar{e}_{p}\gamma_\mu d_{\alpha r})
(d_{\beta s}CiD^\mu_{\sigma \rho} d_{\rho t}).
\end{eqnarray}
We often use the notation $(\Psi C\chi)=\overline{\Psi^C}\chi$ for a bilinear involving charge-conjugated fields to avoid too many indices on the fields. The charge-conjugated field is defined as $\Psi^C=C\bar\Psi^T$ with $(\Psi^C)^C=\Psi$, where the matrix $C$ satisfies the relations $C^T=C^\dagger=-C$ and $C^2=-1$. Note that some operators involving identical fields can vanish in special cases; for instance, with one generation of down-type quarks, both $\mathcal{O}^{prst}_{\bar{L}dddH}$ and $\mathcal{O}^{prst}_{\bar{e}Qdd\tilde{H}}$ vanish since their second bilinear factor vanishes.

We are now in a position to verify the claims in the caption to Table \ref{tab1}. First of all, we prove the equivalence between the original operator $\epsilon_{ij}(L^iC\gamma_\mu e)(\bar{d}\gamma^\mu u)H^j$ and the operator $\mathcal{O}_{\bar{d}LueH}$. In the course of our computation we have made free use of the Fierz identities derived in Refs. \cite{Liao:2012uj,Nieves:2003in} for uncontracted products of bilinears and products of bilinears involving charge-conjugated fields respectively. Some identities are collected in the Appendix. Note that the Fierz identities are basically algebraic identities for gamma matrices though we need here those written for fermion fields. Using the Fierz identity for chiral fields,
\begin{align}
\label{fierz1}
(\overline{\Psi_{1L}^C}\gamma_\mu \Psi_{2R})( \overline{\Psi_{3R}} \gamma^\mu \Psi_{4R})
&=2(\overline{\Psi_{3R}}\Psi_{1L})(\overline{\Psi_{4R}^C}\Psi_{2R}),
\end{align}
where anticommutativity of fermion fields has been taken into account, we have indeed
\begin{equation}
\label{fierz4}
 \epsilon_{ij}(L^iC\gamma_\mu e)(\bar{d}\gamma^\mu u)H^j
 =2\epsilon_{ij}(\bar{d}L^i)(uCe)H^j
 =2\mathcal{O}_{\bar{d}LueH}.
\end{equation}

Now we demonstrate that the operators $\mathcal{O}^{(2)}_{\bar{d}uLLD}$ and $\mathcal{O}_{\bar{L}dQdD}$ can be expressed in terms of other operators and can thus be dropped as redundant operators. Writing $\sigma^{\mu\nu}=i\gamma^\mu\gamma^\nu-ig^{\mu\nu}$ and employing Eq. (\ref{fierz4}) and the EoM (\ref{eom1}), we obtain
\begin{eqnarray}
\label{relation1}
\nonumber
\mathcal{O}^{(2)prst}_{\bar{d}uLLD}&=&
\epsilon_{ij}(\bar{d}_p\gamma_\mu u_r)(L^i_sC\sigma^{\mu\nu}D_\nu L^j_t)
\\
\nonumber
&=&\epsilon_{ij}(\bar{d}_p\gamma_\mu u_r)(L^i_sC\gamma^\mu\gamma^\nu iD_\nu L^j_t)-\epsilon_{ij}(\bar{d}_p\gamma_\mu u_r)(L^i_sCiD^\mu L^j_t)
\\
\nonumber
&=&(Y_e)_{tu}\epsilon_{ij}(\bar{d}_p\gamma_\mu u_r)(L^i_sC\gamma^\mu e_u)H^j-\mathcal{O}^{prst}_{\bar{d}uLLD}
\\
&=&2(Y_e)_{tu}\mathcal{O}^{psru}_{\bar{d}LueH}-\mathcal{O}^{prst}_{\bar{d}uLLD},
\end{eqnarray}
where we have attached the flavor indices but suppressed the color indices. We can thus remove $\mathcal{O}^{(2)}_{\bar{d}uLLD}$ in favor of $\mathcal{O}_{\bar{d}LueH}$ and $\mathcal{O}_{\bar{d}uLLD}$. To show that the operators $\mathcal{O}_{\bar{L}dQdD}$, $\mathcal{O}_{\bar{L}QddD}$, and $\mathcal{O}_{\bar{L}dQQ\tilde H}$ are not independent, we employ the Fierz identity,
\begin{align}
\label{fierz2}
(\overline{\Psi_{1L}}\gamma_\mu\Psi_{2L})(\Psi_{3R}C\Psi_{4R})
&=( \overline{\Psi_{1L}} \Psi_{3R})( \Psi_{2L}C\gamma_\mu\Psi_{4R})
+( \overline{\Psi_{1L}} \Psi_{4R})(\Psi_{2L}C\gamma_\mu \Psi_{3R}).
\end{align}
Replacing $(\Psi_{1L}, \Psi_{2L}, \Psi_{3R}, \Psi_{4R})$ by $(L_{ip}, Q_{i\alpha r}, d_{\beta s}, iD^\mu_{\sigma \rho}d_{\rho s})$ and applying the EoM (\ref{eom2}), the operator $\mathcal{O}^{prst}_{\bar{L}QddD}$ can be reduced as follows:
\begin{eqnarray}
\label{relation2}
\nonumber
\mathcal{O}^{prst}_{\bar{L}QddD}&=&
\epsilon_{\alpha \beta\sigma}\delta_{ij}(\bar{L}_{ip}\gamma_\mu Q_{j\alpha r})(d_{\beta s}
CiD^\mu_{\sigma \rho} d_{\rho t})
\\
\nonumber
&=&\epsilon_{\alpha \beta\sigma}\delta_{ij}\Big((\bar{L}_{ip}d_{\beta s})(Q_{j\alpha r}
Ci\gamma_\mu D^\mu_{\sigma \rho} d_{\rho t})
+ (\bar{L}_{ip}iD^\mu_{\sigma \rho} d_{\rho t})(Q_{j\alpha r}C\gamma_\mu d_{\beta s})\Big)
\\
\nonumber
&=&(Y^\dagger_d)_{tu}\epsilon_{\alpha \beta\sigma}\delta_{ij}\delta_{kl}(\bar{L}_{ip}d_{\beta s})
(Q_{j\alpha r}CQ_{k\sigma u})H^*_l+\mathcal{O}^{ptrs}_{\bar{L}dQdD}
\\
&=&(Y^\dagger_d)_{tu}\mathcal{O}^{psru}_{\bar{L}dQQ\tilde{H}}+\mathcal{O}^{ptrs}_{\bar{L}dQdD}.
\end{eqnarray}
The second equality in the above can also be established by first employing a pure algebraic Fierz identity
\begin{eqnarray}
(\gamma_\mu P_\pm)_{\rho\sigma}(P_\mp)_{\alpha\beta}
&=&(P_\mp)_{\rho\beta}(\gamma_\mu P_\pm)_{\alpha\sigma}
+(P_\mp C^{-1})_{\rho\alpha}(C\gamma_\mu P_\mp)_{\sigma\beta},
\end{eqnarray}
where $P_\pm$ projects out the right- and left-handed chirality respectively, and then attaching the spinor components of the above fields. Note that a spinor being acted upon beforehand by a covariant derivative or gamma matrices does not hinder this application. One can confirm the above identity by using, e.g., Eqs. (27) and (30) in \cite{Liao:2012uj}, and multiplying the $C$ matrix judiciously. Equation (\ref{relation2}) implies that we can remove $\mathcal{O}_{\bar{L}dQdD}$ as redundant as shown in Table \ref{tab1}.

In summary, there are 18 independent dim-7 operators, out of which 6 are in the set of $-\Delta L=\Delta B=1$ and 12 in the set of $\Delta L=2,~\Delta B=0$. We thus have one less operator in each set than Ref.~\cite{Lehman:2014jma}, and both redundant operators are in the class $\psi^4D$ in Table \ref{tab1}. The number of operators has also been counted previously in \cite{Lehman:2015coa} by Hilbert series methods and in \cite{Henning:2015alf} by conformal algebra. Those papers count independent operators that also take into account independent flavor indices for $n$ generations of fermions. We here summarize the differences. While Ref.~\cite{Henning:2015alf} only counts the total number of operators in each class, Ref.~\cite{Lehman:2015coa} counts each type of operators in each class (except for the class $\psi^2H^2D^2$ and part of the class $\psi^4H$). The difference arises in the class $\psi^2H^2D^2$, as already pointed out in \cite{Henning:2015alf}: Ref.~\cite{Henning:2015alf} finds $n(n+1)$ operators in total while Ref.~\cite{Lehman:2015coa} finds $n(n+3)/2$. (We do not include factor of two accounting for Hermitian conjugate of each operator since all dim-7 operators are non-Hermitian.) Using our basis of 18 operators, we have also counted independent operators that take into account flavor indices. We have managed to do so by exhausting all flavor symmetries for each operator, with the simplest ones shown in Eqs. (\ref{Orelation}), that can be employed to remove redundancy. We confirmed separate counts for each operator in \cite{Lehman:2015coa} with the exception for the class $\psi^2H^2D^2$: we found the same number $n(n+1)/2$ of $\calO_{LHD1}$ and $\calO_{LHD2}$, thus confirming the total number in \cite{Henning:2015alf}. (In passing, we note a typo in \cite{Lehman:2015coa} concerning the number of $\calO_{\bar eLLLH}$, with the correct number being $n^2(2n^2+1)/3$.) Had the two redundant operators in Table \ref{tab1} not been deleted, the number of operators in the class $\psi^4D$ would not match with Refs.~\cite{Lehman:2015coa,Henning:2015alf}.

\section{Renormalization group equations for Wilson coefficients}
\label{rge}
The effective Lagrangian for dim-7 operators is written symbolically as
\begin{equation}
\mathcal{L}_7=\sum_i C_i\mathcal{O}_i + \textrm{h.c.},
\end{equation}
where $C_i$ is the Wilson coefficient associated with the operator $\mathcal{O}_i$. The index $i$ enumerates all 18 operators shown in Table \ref{tab1} which are all non-Hermitian, and the sum over $i$ also covers the flavor indices of quark and lepton fields. To study the effects of the above interactions in low energy processes, it is necessary to run the operators from the high scale at which they are generated to the low scale at which their matrix elements are evaluated. The running effect is governed by RGEs and is incorporated in their Wilson coefficients. In this work, we study the RGEs for the subset of operators in Eq. (\ref{eq_BL}) at one-loop level. This is self-consistent since those six operators violate baryon number and do not mix at one loop with the remaining twelve operators which conserve baryon number.

The renormalization group equations for the Wilson coefficients $C_i$ are
\begin{equation}
\dot{C}_i\equiv16\pi^2\mu\frac{d C_i}{d\mu}=\sum^6_{j=1}\gamma_{ij}C_j,
\end{equation}
where $\mu$ is the renormalization scale, $\gamma_{ij}$ is the $6\times 6$ anomalous dimension matrix, and $ij$ enumerate the six operators in Eq. (\ref{eq_BL}). We will work with dimensional regularization in $D=4-2\epsilon$ dimensions and adopt the minimal subtraction ($\overline{\MS}$) scheme. We compute in general $R_\xi$ gauge with three separate gauge parameters $\xi_{1,2,3}$ for three gauge fields. The complete cancellation of all $\xi_{1,2,3}$ dependence in the $\gamma$ matrix serves as a strong check on our result. Before presenting our results, we notice some symmetries in flavor indices. The operators have the following relations,
\begin{eqnarray}
\calO^{p r s t}_{\bar{L}dddH}+\calO^{prts}_{\bar{L}dddH}=0,~ \calO^{prst}_{\bar{L}dddH}+\calO^{pstr}_{\bar{L}dddH}+\calO^{ptrs}_{\bar{L}dddH}=0,~
\calO^{prst}_{\bar{e}Qdd\tilde{H}}+\calO^{prts}_{\bar{e}Qdd\tilde{H}}=0,
\label{Orelation}
\end{eqnarray}
where the first and last ones are obvious by inspection and the second one is obtained by further using the last identity in Eq. (\ref{eq_A1}). These relations are helpful to organize our computational results in the standard basis.

\begin{figure}[!htbp]
\begin{center}
\includegraphics[width=0.8\linewidth]{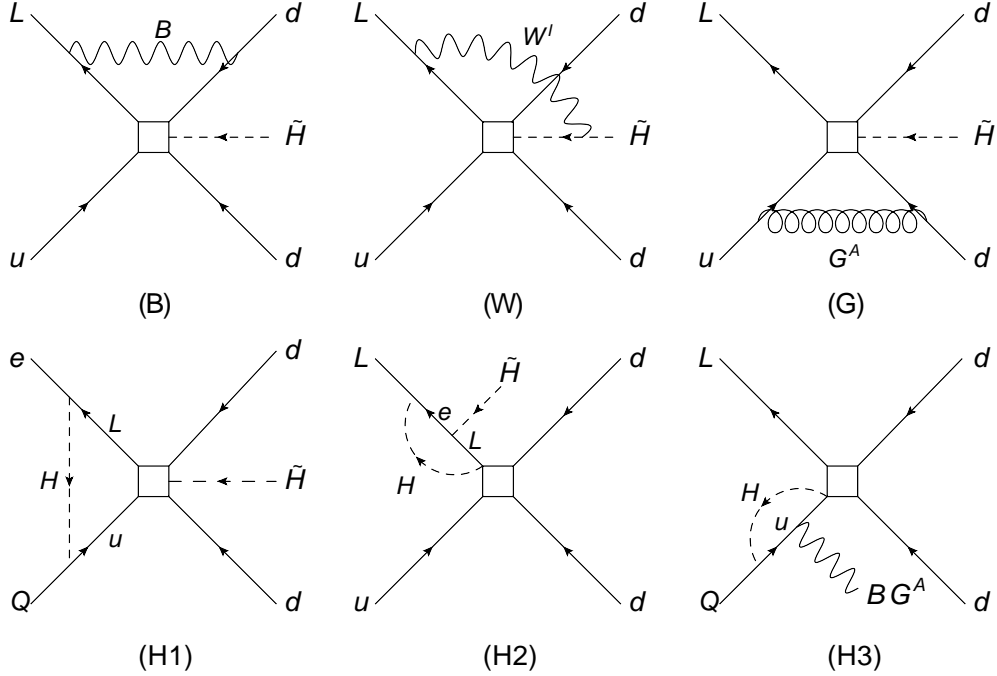}
\end{center}
\caption{Representative one-loop Feynman diagrams with an insertion of the effective interaction $C_{\bar{L}dud\tilde{H}}\mathcal{O}_{\bar{L}dud\tilde{H}}$, shown as a box, from ${\cal L}_7$. They are organized into six categories, (B)--(H3). Categories (B), (W), and (G) stand for the insertion of an internal $B$, $W^I$, and $G^A$ gauge boson propagator, respectively; categories (H1), (H2), and (H3) stand for the insertion of an internal scalar $H$ between two fermions in various ways with the release of a scalar $H$ or gauge field $(B,~W^I,~G^A)$. The total numbers of Feynman diagrams for those six categories in this example are 10 for (B), 1 for (W), 3 for (G), 3 for (H1), 4 for (H2), and 1 for (H3).}
\label{fig1}
\end{figure}

We are now ready to study the one-loop renormalization of the dim-7 interactions ${\cal L}_7$ due to the SM interactions ${\cal L}_4$. We will not present the lengthy computational details; for the purpose of illustration, let us consider the one-loop Feynman diagrams with the insertion of the effective interaction $C_{\bar{L}dud\tilde{H}}\mathcal{O}_{\bar{L}dud\tilde{H}}$. The representative diagrams are shown in Fig. \ref{fig1}, and are classified into six categories from (B) to (H3). The diagrams with the insertion of other three operators involving a Higgs field are similarly classified, but those with the insertion of an operator involving a covariant derivative have more categories. We compute graphs as a contribution to the relevant amplitude. For instance, the first graph in Fig. \ref{fig1} that involves the exchange of a $B$ gauge field between the lepton doublet $L$ (of hypercharge $y_L$) and the singlet $d$ quark (of hypercharge $y_d$) yields a term in the amplitude,
\begin{eqnarray}
\mathcal{M}&=&C^{prst}_{\bar{L}dud\tilde{H}}\epsilon_{\alpha \beta\sigma}\epsilon_{ij}
\mu^{4-D}\!\!\int\frac{d^Dk}{(2\pi)^D}
\left(\bar{L}_{ip}ig_1y_L\gamma_\mu\frac{i}{\slashed{k}_1}\frac{i}{\slashed{k}_2}
ig_1y_d\gamma_\nu d_r\right)\left(u_{\beta s}Cd_{\sigma t}\right)
H^*_j\frac{-i}{k^2}G^{\mu\nu}_{\xi_1}(k),
\end{eqnarray}
where $G^{\mu\nu}_{\xi_1}(k)=g^{\mu\nu}+(\xi_1-1)k^\mu k^\nu/k^2$ and the Higgs field is attached for clarity. For the sake of isolating ultraviolet divergences, $k_{1,2}$ can be identified with $k$. Finishing the above loop integral yields a term that is regarded as a contribution from the effective interaction, $g_1^2/(16\pi^2\epsilon)(\xi_1+3)C^{prst}_{\bar{L}dud\tilde{H}}
\mathcal{O}^{prst}_{\bar{L}dud\tilde{H}}$. After all one-loop diagrams are finished, the relevant counterterms are required to cancel the divergences. Finally, we include field strength renormalization constants and compute the $\gamma$-function in the standard manner.

Using the shortcuts for easier identification of terms,
\begin{eqnarray}
C^{\dots}_{1,2,\dots,6}=C^{\dots}_{\bar{L}dud\tilde{H}},~C^{\dots}_{\bar{L}dddH},~
C^{\dots}_{\bar{e}Qdd\tilde{H}},~C^{\dots}_{\bar{L}dQQ\tilde{H}},~
C^{\dots}_{\bar{L}QddD},~C^{\dots}_{\bar{e}dddD},
\end{eqnarray}
our final result is summarised by the RGEs for the above six Wilson coefficients,
\begin{eqnarray}
\label{o1}
\nonumber
\dot{C}^{prst}_1&=&
+\left(-4g^2_3-\frac{9}{4}g^2_2 -\frac{17}{12}g^2_1+W_H\right)C^{prst}_1
-\frac{10}{3}g^2_1C^{ptsr}_1 -\frac{3}{2}(Y_eY^\dagger_e)_{pv}C^{vrst}_1
\\
\nonumber
&&+3(Y^\dagger_dY_d)_{vr}C^{pvst}_1 +3(Y^\dagger_dY_d)_{vt}C^{prsv}_1 +2(Y^\dagger_uY_u)_{vs}C^{prvt}_1
-2(Y^\dagger_dY_u)_{vs}\left(C^{pvrt}_2+v\leftrightarrow r\right)
\\
\nonumber
&&+4(Y_e)_{pv}(Y_u)_{ws}C^{vwrt}_3
-2\Big((Y_u)_{vs}(Y_d)_{wt}+s\leftrightarrow t\Big)C^{prvw}_4
-\frac{1}{6}\left(11g^2_1+24g^2_3\right)(Y_u)_{vs}C^{pvrt}_5
\\
\nonumber
&&+\frac{1}{6}\left(13g^2_1+48g^2_3\right)(Y_u)_{vs}C^{pvtr}_5
-\frac{3}{2}(Y_d)_{vt}(Y^\dagger_dY_u)_{ws}C^{pvrw}_5
\\
&&
-3(Y_u)_{vs}\Big(( Y_d^\dagger Y_d)_{wt}C^{pvrw}_5-r\leftrightarrow t\Big)
+\frac{3}{2}(Y_e)_{pv}(Y^\dagger_dY_u)_{ws}C^{vrtw}_6,
\\
\label{o2}
\nonumber
\dot{C}^{prst}_2&=&
+\Big(-4g^2_3- \frac{9}{4}g^2_2-\frac{13}{12}g^2_1 +W_H\Big)C^{prst}_2
+\frac{5}{2}(Y_eY^\dagger_e)_{pv}C^{vrst}_2
\\
\nonumber
&&+2\Big((Y^\dagger_dY_d)_{vr}C^{pvst}_2 +(Y^\dagger_dY_d)_{vs}C^{prvt}_2 +(Y^\dagger_dY_d)_{vt}C^{prsv}_2\Big)
\\
\nonumber
&&-\frac{1}{4}\Big[\Big((Y^\dagger_uY_d)_{vs}C^{prvt}_1 +(Y^\dagger_uY_d)_{vr}C^{psvt}_1 +(Y^\dagger_uY_d)_{vs}C^{ptvr}_1\Big)-s\leftrightarrow t\Big]
\\
\nonumber
&& +\Big\{\Big[\Big(\frac{1}{3}(g^2_1-6g^2_3)(Y_d)_{vr}C^{pvst}_5 -\frac{1}{4}g^2_1(Y_d)_{vs}C^{pvrt}_5 -\frac{3}{4}(Y_d)_{vr}(Y^\dagger_dY_d)_{wt}C^{pvsw}_5\Big)
+r\leftrightarrow t\Big]-s\leftrightarrow t\Big\}
\\
&&+\frac{1}{2}(Y_e)_{pv}\Big\{\Big[g_1^2\left(C^{vrst}_6 +r\leftrightarrow s\right) +\frac{3}{4}\Big((Y^\dagger_dY_d)_{wt}\left(C^{vrsw}_6 +r\leftrightarrow s\right) +(Y^\dagger_dY_d)_{wr}C^{vtsw}_6\Big)\Big]-s\leftrightarrow t\Big\},
\\
\label{o3}
\nonumber
\dot{C}^{prst}_3&=&
+\Big(-4g^2_3-\frac{9}{4}g^2_2+\frac{11}{12}g^2_1 +W_H\Big)C^{prst}_3
\\
\nonumber
&&+\Big[\Big((Y^\dagger_eY_e)_{pv}C^{vrst}_3 +\frac{5}{4}(Y_uY^\dagger_u+Y_dY^\dagger_d)_{vr}C^{pvst}_3 +3(Y^\dagger_dY_d)_{vs}C^{prvt}_3 -(Y^\dagger_d)_{wr}(Y_d)_{vs}C^{pvwt}_3\Big) -s\leftrightarrow t\Big]
\\
\nonumber
&&-\frac{1}{2}(Y^\dagger_e)_{pv}\Big[\Big((Y^\dagger_u)_{wr}C^{vtws}_1 +2(Y_d)_{ws}C^{vtwr}_4 +(Y_d)_{wt}C^{vsrw}_4 +3g_1^2C^{vrst}_5
+3(Y_d^\dagger Y_d)_{wt}C^{vrsw}_5\Big) -s\leftrightarrow t\Big]
\\
\nonumber
&&+\frac{1}{4}(g^2_1+12g^2_3)(Y^\dagger_d)_{vr}
\Big[\left(C^{pvst}_6 +C^{psvt}_6+C^{pstv}_6\right)-s\leftrightarrow t\Big]
\\
&&-\frac{3}{4}\Big\{\Big[(Y_d^\dagger Y_d)_{vs}(Y_d^\dagger)_{wr}\left(C^{ptvw}_6-r\leftrightarrow v\right)
+(Y_d^\dagger Y_d)_{ws}(Y_d^\dagger )_{vr}\left(C^{ptvw}_6+2C^{pvtw}_6\right)\Big]
-s\leftrightarrow t \Big\},
\\
\label{o4}
\nonumber
\dot{C}^{prst}_4&=&
+\Big(-4g^2_3-\frac{15}{4}g^2_2-\frac{19}{12}g^2_1 +W_H\Big)C^{prst}_4 -3g^2_2 C^{prts}_4 +3(Y^\dagger_dY_d)_{vr}C^{pvst}_4
\\
\nonumber
&&-\frac{1}{2}(Y_eY^\dagger_e)_{pv}\left(4C^{vrts}_4 -C^{vrst}_4\right) +\Big(2(Y_uY^\dagger_u)_{vt}-(Y_dY^\dagger_d)_{vt}\Big)C^{prvs}_4
\\
\nonumber
&&+\frac{1}{2}\Big(5(Y_uY^\dagger_u)_{vs}+(Y_dY^\dagger_d)_{vs}\Big)C^{prvt}_4 +\frac{1}{2}\Big(5(Y_dY^\dagger_d)_{vt}-3(Y_uY^\dagger_u)_{vt}\Big)C^{prsv}_4
\\
\nonumber
&&-(Y_d)_{wr}\Big((Y^\dagger_d)_{vs}C^{pvwt}_4 +(Y^\dagger_d)_{vt}C^{pvsw}_4\Big) -\Big((Y^\dagger_u)_{vs}(Y^\dagger_d)_{wt}(2C^{prvw}_1+C^{pwvr}_1)+s\leftrightarrow t\Big)
\\
\nonumber
&&-2(Y_e)_{pv}(Y^\dagger_d)_{ws}C^{vtwr}_3 -\frac{1}{6}(g^2_1-24g^2_3)(Y^\dagger_d)_{vt}(C^{psvr}_5+r\leftrightarrow v)-\frac{3}{2}(Y_eY_e^\dagger )_{pv}(Y_d^\dagger)_{ws}C^{vtrw}_5
\\
\nonumber
&&+\frac{3}{2}(Y_d^\dagger Y_d)_{vr}(Y_d^\dagger)_{wt}(C^{psvw}_5+v\leftrightarrow w) +\frac{3}{2}\Big(( Y_uY_u^\dagger)_{vs}(Y_d^\dagger)_{wt}+s\leftrightarrow t\Big)C^{pvrw}_5
\\
&&+\frac{3}{2}(Y_e)_{pv}(Y^\dagger_d)_{ws}(Y^\dagger_d)_{xt} \Big(C^{vxrw}_6+C^{vrwx}_6+C^{vrxw}_6\Big),
\\
\label{o5}
\nonumber
\dot{C}^{prst}_5&=&
+\Big(\frac{5}{9}g^2_1-\frac{4}{3}g^2_3\Big)C^{prst}_5 -\Big(\frac{1}{9}g^2_1-\frac{8}{3}g^2_3\Big)C^{prts}_5 +\frac{1}{2}(Y_eY^\dagger_e)_{pv}C^{vrst}_5 +\frac{1}{2}(Y_uY^\dagger_u+Y_dY^\dagger_d)_{vr}C^{pvst}_5
\\
\nonumber
&&+(Y^\dagger_dY_d)_{vs}C^{prvt}_5 +(Y^\dagger_dY_d)_{vt}C^{prsv}_5 -(Y^\dagger_d)_{wr}\left((Y_d)_{vs}C^{pvwt}_5 +(Y_d)_{vt}C^{pvsw}_5\right)
\\
&&-(Y_e)_{pv}(Y^\dagger_d)_{wr}\left(C^{vwst}_6+C^{vswt}_6+C^{vstw}_6\right),
\end{eqnarray}
\begin{eqnarray}
\label{o6}
\nonumber
\dot{C}^{prst}_6&=&
-\Big(\frac{4}{27}g^2_1+\frac{8}{3}g^2_3\Big)C^{prst}_6 -\Big(\frac{2}{9}g^2_1-\frac{4}{3}g^2_3\Big)\Big(C^{prts}_6+C^{psrt}_6+C^{pstr}_6+C^{ptrs}_6 +C^{ptsr}_6\Big)
\\
&&+(Y^\dagger_eY_e)_{pv}C^{vrst}_6+(Y^\dagger_dY_d)_{vr}C^{pvst}_6
+(Y^\dagger_dY_d)_{vs}C^{prvt}_6+(Y^\dagger_dY_d)_{vt}C^{prsv}_6 -2(Y^\dagger_e)_{pv}(Y_d)_{wr}C^{vwst}_5,
\end{eqnarray}
where $W_H={\Tr}(3Y^\dagger_uY_u+3Y^\dagger_dY_d+Y^\dagger_eY_e)$ arises from the Higgs field wavefunction renormalization constant due to Yukawa interactions. We see from the above results that while operators involving a covariant derivative renormalize those involving a Higgs field the opposite does not occur. This interesting phenomenon is consistent with the nonrenormalization theorem formulated recently in Ref. \cite{Cheung:2015aba}.

\begin{figure}[!htbp]
\begin{center}
\includegraphics[width=0.5\linewidth]{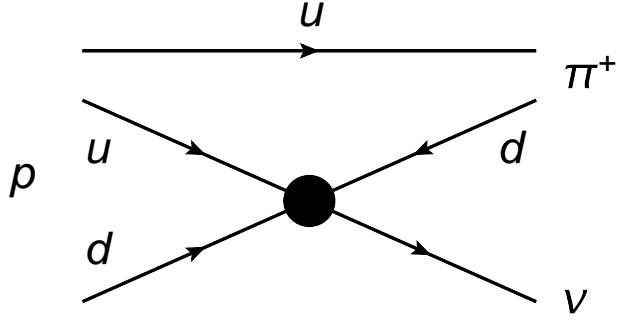}
\end{center}
\caption{Feynman diagram relevant for the decay $p\to\nu\pi^+$ due to dim-7 effective interactions.}
\label{fig2}
\end{figure}

The above dim-7 operators violate both baryon and lepton numbers by one unit but preserve their sum, and would contribute to the rare nucleon decays such as $p\to\nu\pi^+,~\nu K^+$, $n\to e^- \pi^+$, and $p\to e^-\pi^+\pi^+$, and so on. We will not attempt here a complete analysis on this which would involve a sequence of low energy theories below the electroweak scale, but instead illustrate potential impact of the above RGEs by estimating typical running effects. We take the decay $p\to\nu\pi^+$ as an example. We ignore the operators $\calO_{\bar LQddD}$ and $\calO_{\bar edddD}$ which are subleading at low energies, and set the Higgs field $H$ to its vacuum expectation value $v/\sqrt{2}$ for the other four operators. Then, a potentially contributing operator would involve $\bar\nu udd$ simply by charge conservation, as is shown in Fig. \ref{fig2}. An inspection of Eq. (\ref{eq_BL}) shows that only the operators $\calO^{p111}_{\bar Ldud\tilde H}$ and $\calO^{p111}_{\bar LdQQ\tilde H}$ contain such a term, where the superscript $p$ refers to the neutrino flavor and $1$ to the quarks in the first generation.

To go further, we make some plausible approximations. We ignore quark flavor mixing, and drop all Yukawa couplings except for the top quark. Then RGEs (\ref{o1}) and (\ref{o4}) are decoupled,
\begin{eqnarray}
\mu\frac{d}{d\mu}C^{p111}_{\bar{L}dud\tilde{H}}&=&
\frac{1}{4\pi}\left(-4\alpha_3-\frac{9}{4}\alpha_2 -\frac{57}{12}\alpha_1+3\alpha_t\right)C^{p111}_{\bar{L}dud\tilde{H}},
\\
\mu\frac{d}{d\mu}C^{p111}_{\bar{L}dQQ\tilde{H}}
&=&\frac{1}{4\pi}\left(-4\alpha_3-\frac{27}{4}\alpha_2 -\frac{19}{12}\alpha_1+3\alpha_t\right)C^{p111}_{\bar{L}dQQ\tilde{H}},
\end{eqnarray}
where $\alpha_i=g^2_i/(4\pi)~(i=1,2,3)$ and $\alpha_t=Y_t^2/(4\pi)$. The solutions for running from a high energy scale $M\sim 10^{15}~\GeV$ of order grand unification scale to a low energy scale $\mu\sim m_p\sim 1~\GeV$ of the proton mass are
\begin{eqnarray}
C^{p111}_{\bar{L}dud\tilde{H}}(m_p)&=&
\left[\frac{\alpha_3(m_p)}{\alpha_3(M)}\right]^{2/\beta_3} \left[\frac{\alpha_2(M_Z)}{\alpha_2(M)}\right]^{9/(8\beta_2)}
\left[\frac{\alpha_1(M_Z)}{\alpha_1(M)}\right]^{57/(24\beta_1)} (0.787)C^{p111}_{\bar{L}dud\tilde{H}}(M),
\\
C^{p111}_{\bar{L}dQQ\tilde{H}}(m_p)&=&
\left[\frac{\alpha_3(m_p)}{\alpha_3(M)}\right]^{2/\beta_3} \left[\frac{\alpha_2(M_Z)}{\alpha_2(M)}\right]^{27/(8\beta_2)} \left[\frac{\alpha_1(M_Z)}{\alpha_1(M)}\right]^{19/(24\beta_1)} (0.787)C^{p111}_{\bar{L}dQQ\tilde{H}}(M),
\end{eqnarray}
where we have solved numerically the running effect of $\alpha_t$ from $M$ to the electroweak scale of the $Z$-boson mass $M_Z$ (factor $0.787$) using the one-loop $\beta_i$ functions,
\begin{equation}
\beta_3=7, \quad \beta_2=\frac{19}{6}, \quad \beta_1=-\frac{41}{10},
\end{equation}
and the $\overline{\MS}$ values of $\alpha_i~(i=1,2,3,t)$ at $M_Z$ \cite{Mihaila:2012pz},
\begin{align}
\nonumber
\alpha_1(M_Z)=&0.0169225\pm0.0000039, \quad \alpha_2(M_Z)=0.033735\pm0.000020,
\\
\alpha_3(M_Z)=&0.1173\pm0.00069,\quad \alpha_t(M_Z)=0.07514
\end{align}
The overall RGE running results are
\begin{align}
C^{p111}_{\bar{L}dud\tilde{H}}(m_p)=&
(2.034)(1.158)(1.262)(0.787)C^{p111}_{\bar{L}dud\tilde{H}}(M)
=2.34C^{p111}_{\bar{L}dud\tilde{H}}(M),
\\
C^{p111}_{\bar{L}dQQ\tilde{H}}(m_p)=&
(2.034)(1.551)(1.081)(0.787)C^{p111}_{\bar{L}dQQ\tilde{H}}(M)
=2.68C^{p111}_{\bar{L}dQQ\tilde{H}}(M),
\end{align}
where the numerical factors come from the three gauge interactions and top quark Yukawa coupling, respectively. We see that while gauge interactions tend to enhance the effective interactions at low energies, the top quark Yukawa coupling suppresses it, with a balanced enhancement factor of about two.

\section{Conclusion}\label{conclusion}

We have studied dimension-seven operators in the framework of standard model effective field theory. All of these operators violate lepton number conservation. We found that the basis of twenty operators listed in Ref. \cite{Lehman:2014jma} can be further reduced using the equations of motion in the standard model and Fierz identities. The final basis contains twelve operators that conserve baryon number and six operators that break it. We have computed for the first time the anomalous dimension matrix for the latter set of operators by taking into account all interactions in the standard model. We illustrated its possible effect in the rare proton decay $p\to\nu\pi^+$ and found that the renormalization running effect in the relevant Wilson coefficients is about a factor two enhancement from the grand unification scale to the nucleon mass scale.

\vspace{0.5cm}
\noindent %
\section*{Acknowledgement}

This work was supported in part by the Grants No. NSFC-11025525, No. NSFC-11575089 and by the CAS Center for Excellence in Particle Physics (CCEPP). One of us (YL) is grateful to Prof. Xiao-Yuan Li at ITP, CAS, for enlightening discussions. We thank Dr. L. Lehman for electronic communications.

\section*{Appendix: Some useful Fierz identities}

We here summarize Fierz identities for field operators, which are useful for our analysis of operator redundancy and in bringing one-loop renormalized operators back to the standard basis. Our notation for charge conjugation of chiral fields is, $\Psi^C_{L,R}\equiv(\Psi_{L,R})^C$, which has the properties, $\overline{\Psi^C_{L,R}}=(\Psi_{L,R})^TC$ and $\Psi_{L,R}=(\Psi^C_{L,R})^C$. The identities are
\begin{eqnarray}
\label{eq_A1}
\nonumber
( \overline{\Psi_{1L}}\gamma^\mu  \gamma^\nu \Psi_{2R})(\overline{\Psi_{3L}}\gamma_\mu \gamma_\nu \Psi_{4R})
&=&8[(\overline{\Psi_{1L}}\Psi_{2R})(\overline{\Psi_{3L}}\Psi_{4R}) +(\overline{\Psi_{1L}}\Psi_{4R})(\overline{\Psi_{3L}} \Psi_{2R})], \ (L\leftrightarrow R)
\\
\nonumber
( \overline{\Psi_{1L}}\gamma^\mu  \gamma^\nu \Psi_{2R})(\overline{\Psi_{3L}}\gamma_\nu \gamma_\mu \Psi_{4R})
&=&-8(\overline{\Psi_{1L}}\Psi_{4R})(\overline{\Psi_{3L}} \Psi_{2R})
\\
\nonumber
( \overline{\Psi_{1L}}\gamma^\mu  \gamma^\nu \Psi_{2R})(\overline{\Psi_{3R}}\gamma_\mu \gamma_\nu \Psi_{4L})
&=&4(\overline{\Psi_{1L}}\Psi_{2R})(\overline{\Psi_{3R}}\Psi_{4L})
\\
\nonumber
( \overline{\Psi_{1L}}\gamma^\mu  \gamma^\nu \Psi_{2R})(\overline{\Psi_{3R}}\gamma_\nu \gamma_\mu \Psi_{4L})
&=&4(\overline{\Psi_{1L}}\Psi_{2R})(\overline{\Psi_{3R}} \Psi_{4L})
\\
\nonumber
( \overline{\Psi_{1L}} \gamma^\mu \Psi_{2L})( \overline{\Psi_{3L}}  \gamma_\mu \Psi_{4L})
&=&(\overline{\Psi_{1L}} \gamma^\mu \Psi_{4L})(\overline{\Psi_{3L}}\gamma_\mu\Psi_{2L}), \ (L\leftrightarrow R)
\\
\nonumber
( \overline{\Psi_{1L}} \gamma^\mu \Psi_{2L})( \overline{\Psi_{3L}}  \gamma_\mu \Psi_{4L})
&=&2(\overline{\Psi_{1L}} \Psi^C_{3L})(\overline{\Psi^C_{4L}}\Psi_{2L}), \ (L\leftrightarrow R)
\\
\nonumber
( \overline{\Psi_{1L}} \gamma^\mu \Psi_{2L})( \overline{\Psi_{3R}}  \gamma_\mu \Psi_{4R})
&=&-2(\overline{\Psi_{1L}} \Psi_{4R})(\overline{\Psi_{3R}}\Psi_{2L})
\\
\nonumber
( \overline{\Psi_{1L}} \gamma^\mu \Psi_{2L})( \overline{\Psi^C_{3R}}  \Psi_{4R})
&=&( \overline{\Psi_{1L}}  \Psi_{3R})( \overline{\Psi^C_{2L}}  \gamma_\mu \Psi_{4R})+( \overline{\Psi_{1L}}  \Psi_{4R})( \overline{\Psi^C_{2L}}  \gamma_\mu \Psi_{3R})
\\
\nonumber
( \overline{\Psi_{1L}}  \Psi_{2R})( \overline{\Psi^C_{3L}} \gamma^\mu  \Psi_{4R})
&=&( \overline{\Psi_{1L}}  \gamma_\mu \Psi_{3L})( \overline{\Psi^C_{4R}}  \Psi_{2R})-( \overline{\Psi_{1L}}  \Psi_{4R})( \overline{\Psi^C_{3L}}  \gamma_\mu \Psi_{2R})
\\
\nonumber
( \overline{\Psi_{1R}} \gamma^\mu \Psi_{2R})( \overline{\Psi^C_{3R}}  \Psi_{4R})
&=&-( \overline{\Psi_{1R}}   \gamma_\mu \Psi_{3R})( \overline{\Psi^C_{2R}}  \Psi_{4R})-( \overline{\Psi_{1R}} \gamma_\mu \Psi_{4R})( \overline{\Psi^C_{2R}}  \Psi_{3R})
\\
( \overline{\Psi_{1L}}  \Psi_{2R})( \overline{\Psi^C_{3R}}  \Psi_{4R})
&=&-( \overline{\Psi_{1L}}\Psi_{3R})( \overline{\Psi^C_{4R}}  \Psi_{2R})-( \overline{\Psi_{1L}}  \Psi_{4R})( \overline{\Psi^C_{3R}} \Psi_{2R})
\end{eqnarray}
From the basic relations for bilinears involving charge conjugation,
\begin{eqnarray}
\overline{\Psi_{1L}}\gamma^{\mu_1}\gamma^{\mu_2} \ \cdot\cdot\cdot \ \gamma^{\mu_{n-1}}\gamma^{\mu_n}\Psi_{2R}
&=&\overline{\Psi^C_{2R}}\gamma^{\mu_n}\gamma^{\mu_n-1}\ \cdot\cdot\cdot \ \gamma^{\mu_{2}}\gamma^{\mu_1}\Psi^C_{1L}, \textrm{ for $n$ even} \ (L\leftrightarrow R)
\nonumber
\\
\overline{\Psi_{1L}}\gamma^{\mu_1}\gamma^{\mu_2} \ \cdot\cdot\cdot \ \gamma^{\mu_{n-1}}\gamma^{\mu_n} \Psi_{2L}
&=&-\overline{\Psi^C_{2L}}\gamma^{\mu_n}\gamma^{\mu_n-1}\ \cdot\cdot\cdot \ \gamma^{\mu_{2}}\gamma^{\mu_1}\Psi^C_{1L}, \textrm{ for $n$ odd} \ (L\leftrightarrow R)
\label{c3}
\end{eqnarray}
we have the special cases
\begin{eqnarray}
\nonumber
\overline{\Psi_{1L}}\Psi_{2R}&=&\overline{\Psi^C_{2R}}\Psi^C_{1L} \ (L\leftrightarrow R)
\\
\nonumber
\overline{\Psi_{1L}}\gamma^\mu \Psi_{2L}&=&-\overline{\Psi^C_{2L}}\gamma^\mu \Psi^C_{1L} \ (L\leftrightarrow R)
\\
\overline{\Psi_{1L}}\gamma^{\mu}\gamma^{\nu}\Psi_{2R}&=& \overline{\Psi^C_{2R}}\gamma^{\nu}\gamma^{\mu}\Psi^C_{1L} \ (L\leftrightarrow R)
\end{eqnarray}
\noindent %

\end{document}